\documentclass[a4paper,11pt]{article}
\pdfoutput=1 

\usepackage{jcappub} 

\usepackage{graphicx}
\usepackage{siunitx}
\usepackage{physics}

\abstract{
Axion haloscopes search for dark matter axions from the galactic halo, most commonly by measuring a power excess sourced by the axion effective current density. Constraining axion parameters from detection or lack thereof requires estimating the expected signal power. Often, this is done by studying the response of the haloscope to a known, but different, source current density, for example via a reflection measurement. However, only in the special case when both sources induce the same electromagnetic fields, do the quantities derived from a reflection measurement adequately describe the setup during an axion measurement. While this might be valid for the traditional resonant cavity haloscope, new broadband or open designs like dish antennas or dielectric haloscopes cannot make this assumption. A more general relation between axion- and reflection-induced fields is needed. In this article, we use the Lorentz reciprocity theorem to derive an expression for the axion signal power which instead of the unmeasurable axion-induced fields depends on the measurable reflection-induced fields. This entirely circumvents the need to know the response of the haloscope to the unknown axion source. It applies to a wide variety of haloscopes including resonant cavities, dielectric haloscopes, and broadband dish antennas. 
}

\begin{document}

\title{Axion haloscope signal power from reciprocity}

\author{Jacob Egge}
\affiliation{Universität Hamburg,
22761 Hamburg,
Germany
}

\emailAdd{jacob.egge@posteo.de}


\date{\today}


\maketitle
\flushbottom

\section{\label{sec:intro}Introduction}
The composition of dark matter might be the greatest unsolved problem in cosmology. The axion is a particle candidate that has gathered increasing interest over the past years. It was originally introduced via the Peccei-Quinn mechanism \cite{peccei_quinn1977} to solve the so-called strong CP problem: the surprising absence of charge-parity (CP) violations in the strong interaction. Due to its weak coupling to standard model particles, the axion is also a natural dark matter candidate and could thus solve two problems at once. Axion haloscopes aim to detect axions from the galactic dark matter halo (see \cite{semertzidis2022} for a recent overview). The majority of haloscopes rely on the axion-photon interaction to convert dark matter axions to detectable photons
\begin{equation}
    \mathcal{L}_{int} = g_{a\gamma} a \vb*{E}\vdot \vb*{B},
    \label{Eq:lagrangian}
\end{equation}
where the constant $g_{a\gamma}$ describes the coupling strength between axion field $a$ and electromagnetic fields $\vb*{E}$ and $\vb*{B}$. Throughout this article, we will use Lorentz-Heaviside units ($\hbar=c=\epsilon_0=\mu_0=1$). The dark matter axion field $a$ oscillates at an unknown angular frequency $\omega_a$ that to lowest order in $g_{a\gamma}$ depends on axion mass $m_a$ and momentum $k_a$ via the free axion dispersion relation $\omega_a^2=m_a^2+k_a^2$ \cite{millar2017}. The axion mass $m_a$ is still poorly constrained and can provide the correct abundance of dark matter over several orders of magnitude from $\SI{e-6}{}$ to $\SI{e3}{\micro \eV}$\cite{arias2012}. This necessitates different setups optimized for a given mass range. Common to most designs is the employment of strong static external magnetic or, less common, electric fields to which the axion can couple. From the axion-photon interaction \eqref{Eq:lagrangian} one can derive axion-modified Maxwell's equations that are in general nonlinear \cite{sikivie1983}. But since the external fields are much larger than any other fields, axion-modified Maxwell's equations can be linearized to first order in $g_{a\gamma}$ \cite{rodd2021}. In particular, Ampère's law obtains an additional term 
\begin{equation}
        \curl{\vb*{H}} - \vb*{\dot{D}} = \vb*{J}_f + g_{a\gamma} \qty(\dot{a} \vb*{B}_e -\vb*{E}_e \cross \grad{a} ).
        \label{Eq:ampere}
\end{equation}
Here, the magnetic field $\vb*{H}$, displacement field $\vb*{D}$, and free current density $\vb*{J}_f$ are purely oscillatory and do not include contributions from the static external fields, $\vb*{B}_e$ and $\vb*{E}_e$. The additional term acts like an effective current density 
\begin{equation}
\vb*{J}_a \equiv g_{a\gamma} \qty(\dot{a} \vb*{B}_e -\vb*{E}_e \cross \grad{a}), 
\label{Eq:axion_current}
\end{equation}
that sources electromagnetic fields $\vb*{E}_a, \vb*{H}_a$ which we will refer to as axion-induced fields. Most haloscopes then try to detect axions by detecting a power excess coming from the axion-induced fields. 

Whether a haloscope finds the axion or not, it is important to constrain its parameters from a detection or lack thereof. Fundamentally, this requires an estimate of the axion signal power from an unknown or absent $\vb*{J}_a$. Experimentally, it is very challenging to set up a known current density in a vacuum as a genuine copy of $\vb*{J}_a$ to directly calibrate the haloscope's response. Instead, haloscopes rely on simulations and complementary measurements where the response to a different source current density is studied. A typical measurement is a reflection measurement where a signal generator (typically from a vector network analyzer) sets up an oscillating current density $\vb*{J}_R$ that then sources the reflection-induced fields $\vb*{E}_R, \vb*{H}_R$. From a reflection measurement, one might then characterize some important quantities of the setup that parameterize the expected axion signal power, for example, the quality factor of a cavity. However, the reflection-induced fields are generally not the same as the axion-induced fields since the respective sources are different. And thus the measured quantities of the setup are fundamentally functions of $\vb*{J}_R$ and not of $\vb*{J}_a$.

This is a central issue haloscopes face and the one we try to address in this article. How can we be sure that the quantities we derive from a reflection measurement also adequately describe the setup during an axion measurement? For closed and resonant systems like the traditional cavity haloscope, the solution can be trivial. If the cavity modes are well-separated, one can confidently assume that in both cases the same resonant mode is excited and so axion-induced and reflection-induced fields are the same, up to normalization. For open, overmoded, or broadband setups this is not possible anymore and we need a more general way of relating axion-induced and reflection-induced fields. The Lorentz reciprocity theorem does exactly this: It relates two different time-harmonic sources and the resulting electromagnetic fields. Reciprocity only applies to linear systems so in addition to linearizing the axion-modified Maxwell's equations we also require linear and time-independent media described by permittivity $\epsilon$, permeability $\mu$, and conductivity $\sigma$ via the usual constitutive relations,
\begin{equation}
    \begin{aligned}
        \vb*{D} &= \epsilon \vb*{E}\\
        \vb*{B} &= \mu \vb*{H}\\
        \vb*{J} &= \sigma \vb*{E}.
    \end{aligned}
    \label{Eq:constitutive}
\end{equation}
For anisotropic materials, the respective tensors need to be symmetric. With these requirements, the Lorentz reciprocity theorem takes on the familiar classical form. In \cite{schulz2018}, the Lorentz reciprocity theorem has been used to calculate the power emitted by a source in an arbitrary optical system in the context of the Purcell effect. The scenario is highly analogous to a general haloscope setup but has gone largely unnoticed in the axion community. 

It is the intention of this article to provide a general description of axion haloscopes that extends beyond the traditional cavity approach. In fact, most of our results will be applicable to arbitrary geometries. We start by describing a general haloscope setup in section \ref{sec:axion_reciprocity} and proceed to derive a simple expression for the axion signal power that no longer depends on the axion-induced fields but on the reflection-induced fields that we can in principle measure. We then apply our results to dish antennas and dielectric haloscopes in section \ref{sec:applications}. In section \ref{sec:num_validation} we validate our results with numerical simulations. Before concluding, we briefly discuss some experimental considerations in section \ref{sec:exp_consideration}.

\section{\label{sec:axion_reciprocity}Haloscope signal power}

\subsection{\label{sec:setup}General haloscope setup}

\begin{figure}
    \centering
    \includegraphics[width=0.8\textwidth]{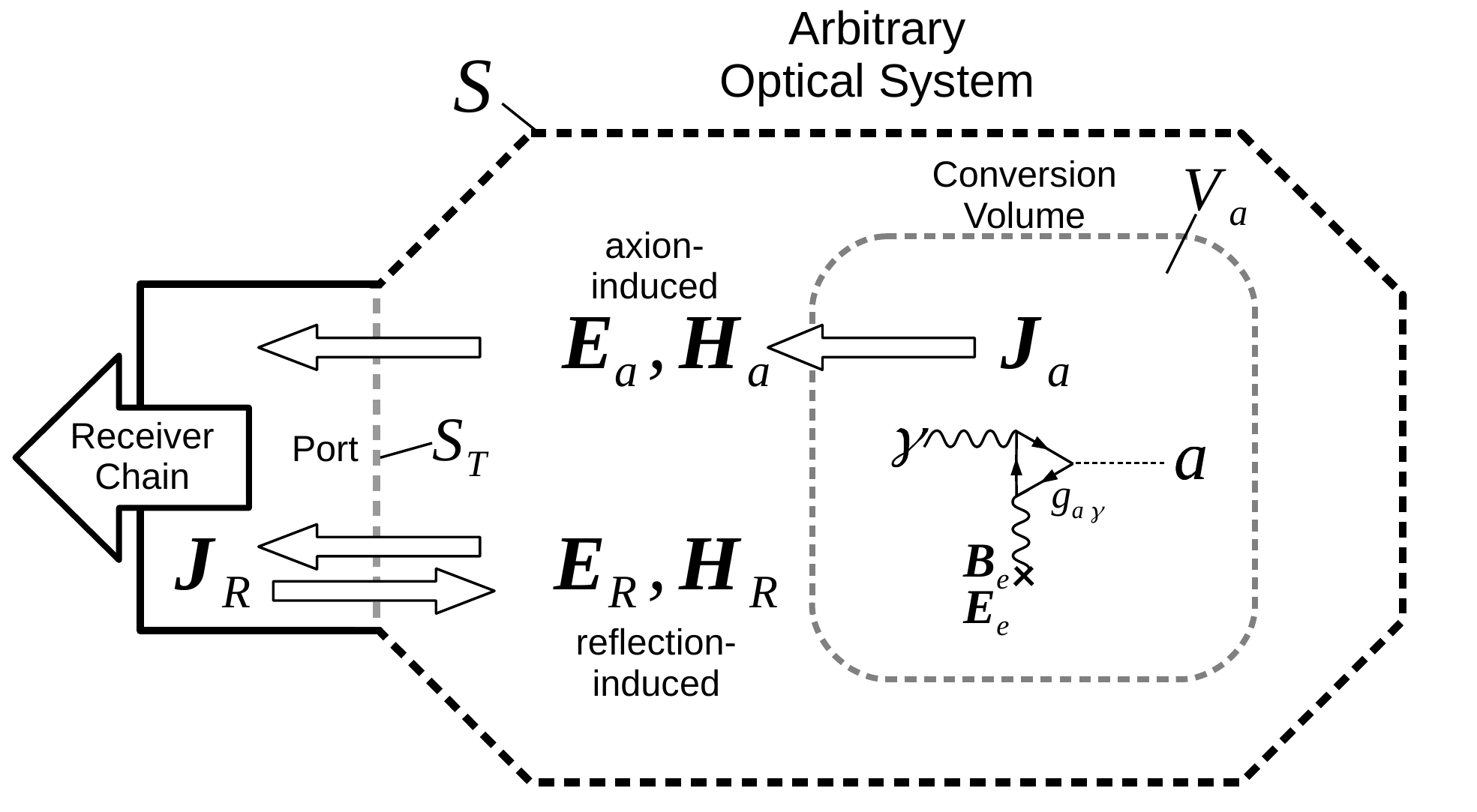}
    \caption{Setup of a general haloscope with two reciprocal modes of operation. Top: the axion current density $\vb*{J}_a$ sources electromagnetic fields $\vb*{E}_a,\vb*{H}_a$ inside the optical system that is coupled to a receiver via a port. Bottom: an external current density $\vb*{J}_R$ excites the reflection-induced fields $\vb*{E}_R,\vb*{H}_R$ from the outside.}
    \label{fig:general_haloscope_setup}
\end{figure}

We split our general haloscope setup into two parts. An optical system that converts axions to photons and a receiver chain capable of detecting the converted photons. In this article, we will focus on the optical system and assume that the receiver chain is well-understood so that from the axion-induced power output of the optical system one can calculate the total signal power for the full haloscope. Figure \ref{fig:general_haloscope_setup} shows a sketch of an arbitrary optical system. It is bound by a surface $S$ inside which we allow an arbitrary geometry. At $S$ we will later demand either a radiation boundary condition (free space) or impedance boundary condition (lossy conductor) allowing for both open and closed systems. Not all of the optical system is necessarily subject to external magnetic or electric fields and $\vb*{J}_a$ is confined to a finite conversion volume $V_a$ in the sense that its contributions to $\vb*{E}_a,\vb*{H}_a$ from outside $V_a$ are negligible. The receiver chain is connected to the optical system via a lossless, single-mode waveguide or transmission line. This allows the definition of a single-mode port with reference plane $S_T \subset S$. Any passive and linear elements between the receiver chain and the optical system can be redefined to be part of the optical system by simply moving the reference plane $S_T$ towards the receiver chain. To truly decouple the optical system from the receiver chain, we will assume that at $S_T$ the axion-induced fields travel only towards the receiver chain as indicated by the arrows in figure \ref{fig:general_haloscope_setup}. This can be achieved for example by reflectionless matching between the optical system and receiver chain or by employing circulators.\footnote{While circulators violate reciprocity, they need not be part of the optical system as we define it and thus do not affect our results.} From now on we switch to time-harmonic notation for all oscillating fields with time-dependence $e^{-i \omega_a t}$. The cycle-averaged axion signal power seen by the receiver chain is simply the time-averaged Poynting vector integrated over the port with the surface normal vector $\vb*{n}$ pointing towards the receiver chain
\begin{equation}
    P_{\mathrm{sig}} = \frac{1}{2} \mathrm{Re}\int_{S_T} \dd{A} \vb*{n}\vdot\left(\vb*{E}_a \cross \vb*{H}_a^*\right). 
    \label{Eq:power_ax_poynting}
\end{equation}
The axion signal power $P_{\mathrm{sig}}$ can then be further parameterized into useful quantities like quality factor, coupling coefficient, boost factor, etc., depending on the setup. But all derived quantities will also be functions of $\vb*{E}_a,\vb*{H}_a$. In an attempt to constrain these quantities from a reflection measurement, a current density $\vb*{J}_R$ outside of the optical system excites the reflection-induced fields $\vb*{E}_R,\vb*{H}_R$. However, only in the trivial case when axion-induced and reflection-induced fields are the same (up to normalization), can we assume that the electromagnetic properties we measure also apply to the axion-induced fields. If this is not the case, we need a more general way to relate these two scenarios before we can put any constraints on $P_{\mathrm{sig}}$ from a reflection measurement. 


\subsection{\label{sec:signal_power} Reciprocity approach}
The Lorentz reciprocity theorem relates the electromagnetic fields of two different sources. In our case these are $\vb*{E}_a,\vb*{H}_a$ from $\vb*{J}_a$ and $\vb*{E}_R,\vb*{H}_R$ from $\vb*{J}_R$. Since $\vb*{J}_R \gg \vb*{J}_a$, the reflection-induced fields $\vb*{E}_R,\vb*{H}_R$ essentially follow unmodified Maxwell's equations while the axion-induced fields $\vb*{E}_a,\vb*{H}_a$ follow linearized axion modified Maxwell's equations. In this case, the derivation of Lorentz reciprocity follows the classical one \cite{pozar2005} and one finds
\begin{equation}
    \int_V \dd{V} \left( \vb*{E}_a \vdot \vb*{J}_R - \vb*{E}_R \vdot \vb*{J}_a \right) = \oint_{\partial V} \dd{A} \vb*{n}\vdot\left(\vb*{E}_R \cross \vb*{H}_a - \vb*{E}_a \cross \vb*{H}_R\right),
\label{Eq:reciprocity_general}
\end{equation}
for arbitrary volume $V$ and its boundary $\partial V$. We choose $\partial V=S$ and thereby exclude $\vb*{J}_R$. We can thus drop the first integrand on the left-hand side of \eqref{Eq:reciprocity_general} and reduce the integration volume to conversion volume $V_a$ where $\vb*{J}_a$ is non-negligible. The two integrands on the right-hand side of \eqref{Eq:reciprocity_general} cancel under either an impedance boundary condition or radiation condition (or a combination of both) at $S$ \cite{gronwald2005}. This means we can allow both a lossy conductor or free space without any incoming radiation at $S$. Only at $S_T$, we cannot apply either boundary condition since we also have incoming radiation from the reflection measurement. Equation \eqref{Eq:reciprocity_general} then reduces to 
\begin{equation}
    -\int_{V_a} \dd{V} \vb*{E}_R \vdot \vb*{J}_a =   \int_{S_T} \dd{A} \vb*{n}\vdot\left(\vb*{E}_R \cross \vb*{H}_a - \vb*{E}_a \cross \vb*{H}_R\right).
    \label{Eq:reciprocity_select}
\end{equation}
The fields at $S_T$ can be expanded into right (+) and left (-) propagating eigenmodes \cite{SnyderLove:1983} with
\begin{equation}
\begin{aligned}
    \vb*{E}_{a,R} &= a^{+}_{a,R} \vb*{e}^{+} + a^{-}_{a,R} \vb*{e}^{-} \\  
    \vb*{H}_{a,R} &= a^{+}_{a,R} \vb*{h}^{+} + a^{-}_{a,R} \vb*{h}^{-} \\
    \int_{S_T} \dd{A} \vb*{n} &\vdot \left(\vb*{e}^{+}\cross\vb*{h}^{*+}\right)=-1\\
    \vb*{e}^{+} &= \vb*{e}^{*-} \, , \, \vb*{h}^{+} = -\vb*{h}^{*-} .
\end{aligned}
\label{Eq:modes_at_port}
\end{equation}
This applies only to the port as we assumed a lossless single-mode waveguide or transmission line connecting the optical system and receiver chain. The fields inside the optical system are still completely arbitrary and do not have to allow any modal decomposition. Inserting \eqref{Eq:modes_at_port} into \eqref{Eq:reciprocity_select} leads to
\begin{equation}
    \int_{S_T} \dd{A} \vb*{n}\vdot\left(\vb*{E}_R\cross \vb*{H}_a - \vb*{E}_a \cross \vb*{H}_R\right) = 2\left(a_a^{-}a_R^{+} - a_a^{+}a_R^{-}\right).
\end{equation}
Since we assumed that the axion-induced fields at $S_T$ propagate only towards the receiver chain, we can set $a_a^{+}=0$.  Taking the absolute square of equation \eqref{Eq:reciprocity_select} then results in
\begin{equation}
\abs{\int_{V_a} \dd{V} \vb*{E}_R \vdot \vb*{J}_a}^2= 4\abs{a_a^{-}}^2 \abs{a_R^{+}}^2.
\label{Eq:reaction_modal_coeff}
\end{equation}
The same modal decomposition can be applied to equation \eqref{Eq:power_ax_poynting} yielding $P_{\mathrm{sig}} = 1/2\abs{a_a^{-}}^2$. Analogously, we define the cycle averaged incident power $P_{\mathrm{in}} \equiv 1/2\abs{a_R^{+}}^2$ that excites the optical system in a reflection measurement. From \eqref{Eq:reaction_modal_coeff} and inserting \eqref{Eq:axion_current}, we can then express the axion signal power as
\begin{equation}
    P_{\mathrm{sig}} = \frac{g_{a\gamma}^2}{16P_{\mathrm{in}}}\abs{\int_{V_a} \dd{V} \vb*{E}_R \vdot  \left(\dot{a} \vb*{B}_e -\vb*{E}_e \cross \grad{a} \right)}^2.
    \label{Eq:power_ax_reciprocity}
\end{equation}
Equation \eqref{Eq:power_ax_reciprocity} is the main result of this article. The axion signal power $P_{\mathrm{sig}}$ now only depends on measurable quantities controlled by the experiment ($P_{\mathrm{in}}$, $\vb*{E}_R$, $\vb*{B}_e$, $\vb*{E}_e$) and the axion parameters we want to constrain ($g_{a\gamma}$, $a$). Crucially, we no longer need to know how the haloscope reacts to an unknown $\vb*{J}_a$. Instead, we need to know how the haloscope reacts to a reflection measurement. While this might still be a challenging task it is at least not fundamentally impossible anymore. Note that for a linear system $\abs{\vb*{E}_R}\propto \sqrt{P_{\mathrm{in}}}$ so that $P_{\mathrm{sig}}$ is actually independent of $P_{\mathrm{in}}$, as it should be. Another important point is that we only need to know the fields inside the conversion volume and not outside. Any structure outside of the conversion volume can only contribute indirectly to $P_{\mathrm{sig}}$ by modifying $\vb*{E}_R$. Equation \eqref{Eq:power_ax_reciprocity} applies to a wide range of possible haloscope setups. For cavity experiments, it reduces to the well-established expression in terms of form and quality factor (cf. appendix \ref{app:cavity_power}). The derivation presented here can be extended to hidden photons with an important caveat: The hidden photon current density is practically infinite in extent and cannot be confined to the volume $V_a$. Instead, $V_a$ must be defined as the extent of $\vb*{E}_R$ which poses a challenge for open optical systems. 


\section{\label{sec:applications}Applications}
We now apply our main result to more specific cases, namely a simple dish antenna and dielectric haloscope. Before we do so, we first bring equation \eqref{Eq:power_ax_reciprocity} into a more suitable form by considering arbitrarily shaped interfaces inside a homogeneous magnetic field. 


\subsection{Arbitrary magnetized interfaces}
A homogeneous external magnetic field $\vb*{B}_e$ is central to most haloscope designs. In the zero velocity limit of cold dark matter, $k_a \rightarrow 0$, the axion field $a$ is also homogeneous and we can write $a = a_0 e^{-i\omega_a t}$. Let us assume that permeability $\mu$, permittivity $\epsilon$, and conductivity $\sigma$ are each piecewise constant for each component inside the conversion volume $V_a$. For brevity, we will use the complex permittivity $\epsilon \rightarrow \epsilon + i\sigma/\omega_a$ to refer to dielectrics and conductors alike. We split $V_a$ into subvolumes $V_k$ for each homogeneous component. Using equation \eqref{Eq:ampere} and a vector calculus identity, we can write for each $V_k$ 
\begin{equation}
        i \omega_a \int_{V_k} \dd{V} \vb*{E}_R \vdot \vb*{B}_e  =  \oint_{\partial V_k} \dd{A}\vb*{n} \vdot \left(\frac{\vb*{H}_R \cross  \vb*{B}_e}{\epsilon}\right) + \int_{V_k} \dd{V} \frac{\curl{\vb*{B}_e}}{\epsilon}\vdot \vb*{H}_R. 
\label{Eq:ampere2}
\end{equation}
The second term on the RHS can be neglected as long as $\vb*{B}_e$ is homogeneous where $\vb*{H}_R$ is large and only drops off to zero where $\vb*{H}_R$ is small. Experimentally this is often achieved by having the optical system in a cryostat inside a magnet bore. The boundary $\partial V_k$ of each subvolume is either part of an interface, denoted as $S_i$, between two components or part of the outer boundary $\partial V_a$. On each interface, the surface integral on the RHS of \eqref{Eq:ampere2} will be evaluated twice from opposite sides. One thus finds\footnote{The derivation here follows the "magic formula" from the discontinuous Galerkin method. For a digestible derivation, visit https://solmaz.io/notes/discontinuous-divergence-theorem/}  
\begin{equation}
        i \omega_a \int_{V_a} \dd{V} \vb*{E}_R \vdot \vb*{B}_e  = \oint_{\partial V_a} \dd{A} \vb*{n} \vdot \left(\frac{\vb*{H}_R \cross  \vb*{B}_e}{\epsilon}\right) + \sum_i \int_{S_i} \dd{A} \vb*{n} \vdot \left[\left[\frac{\vb*{H}_R \cross  \vb*{B}_e}{\epsilon}\right]\right],
\label{Eq:ampere4}
\end{equation}
where the double square brackets $\left[\left[\vdot\right]\right]$ is the jump operator. For an interface $S_i$ between two components $j$ and $k$ and surface normal vector pointing from $j$ to $k$, the jump operator acting on a vector field  $\vb*{v}$ is defined as
\begin{equation}
    \left[\left[\vb*{v}\right]\right] = \vb*{v}_j - \vb*{v}_k,
\end{equation}
where $\vb*{v}_{j,k}$ is the value of $\vb*{v}$ on either side of the interface. Per definition, $\vb*{B}_e$ is negligible on $\partial V_a$ and we can neglect the first term on the RHS of \eqref{Eq:ampere4}. With this we can now express $P_{\mathrm{sig}}$ in terms of the reflection-induced fields on each magnetized interface:
\begin{equation}
    P_{\mathrm{sig}} = \frac{g_{a\gamma}^2  \abs{a_0}^2}{16 P_{\mathrm{in}}}\abs{\sum_i \int_{S_i} \dd{A} \vb*{n} \vdot \left[\left[\frac{\vb*{H}_R \cross  \vb*{B}_e}{\epsilon}\right]\right]}^2.
    \label{Eq:power_ax_surf_int}
\end{equation}
For non-magnetic materials, $\mu = 1$, $\vb*{B}_e$ is continuous across all interfaces and can be pulled out of the jump operator. For lossless dielectrics, $\vb*{n}\cross\vb*{H}_R$ is also continuous and the jump operator only acts on the inverse permittivity. Equation \eqref{Eq:power_ax_surf_int} shows that we need to constrain $\vb*{H}_R$ only on magnetized interfaces inside the conversion volume that can actually contribute. For example, interfaces whose surface normal vector $\vb*{n}$ points along the external magnetic field do not contribute since the triple product would vanish. They only contribute indirectly by modifying $\vb*{H}_R$ on the interfaces that count. The same holds true for interfaces where the surface integral vanishes due to symmetry. This is a big simplification from equation \eqref{Eq:power_ax_reciprocity} where $\vb*{E}_R$ needs to be known throughout the conversion volume. From equation \eqref{Eq:power_ax_surf_int} we can also identify different strategies to maximize our signal power. We can try to maximize each summand by either increasing $\vb*{B}_e$ or by resonance effects that increase $\vb*{H}_R$ relative to $P_{\mathrm{in}}$. Additionally, we can try to maximize the full sum by adding constructively the contributions of each interface. This is already reminiscent of the working principle of a dielectric haloscope where each interface leads to axion-induced emissions that can constructively interfere and form resonances between interfaces. So far we have kept the geometry fully general. Next, we will choose the specific geometry of a simple dish antenna and dielectric haloscope.

\subsection{Dish antenna}

\begin{figure}
    \centering
    \includegraphics[width=0.8\textwidth]{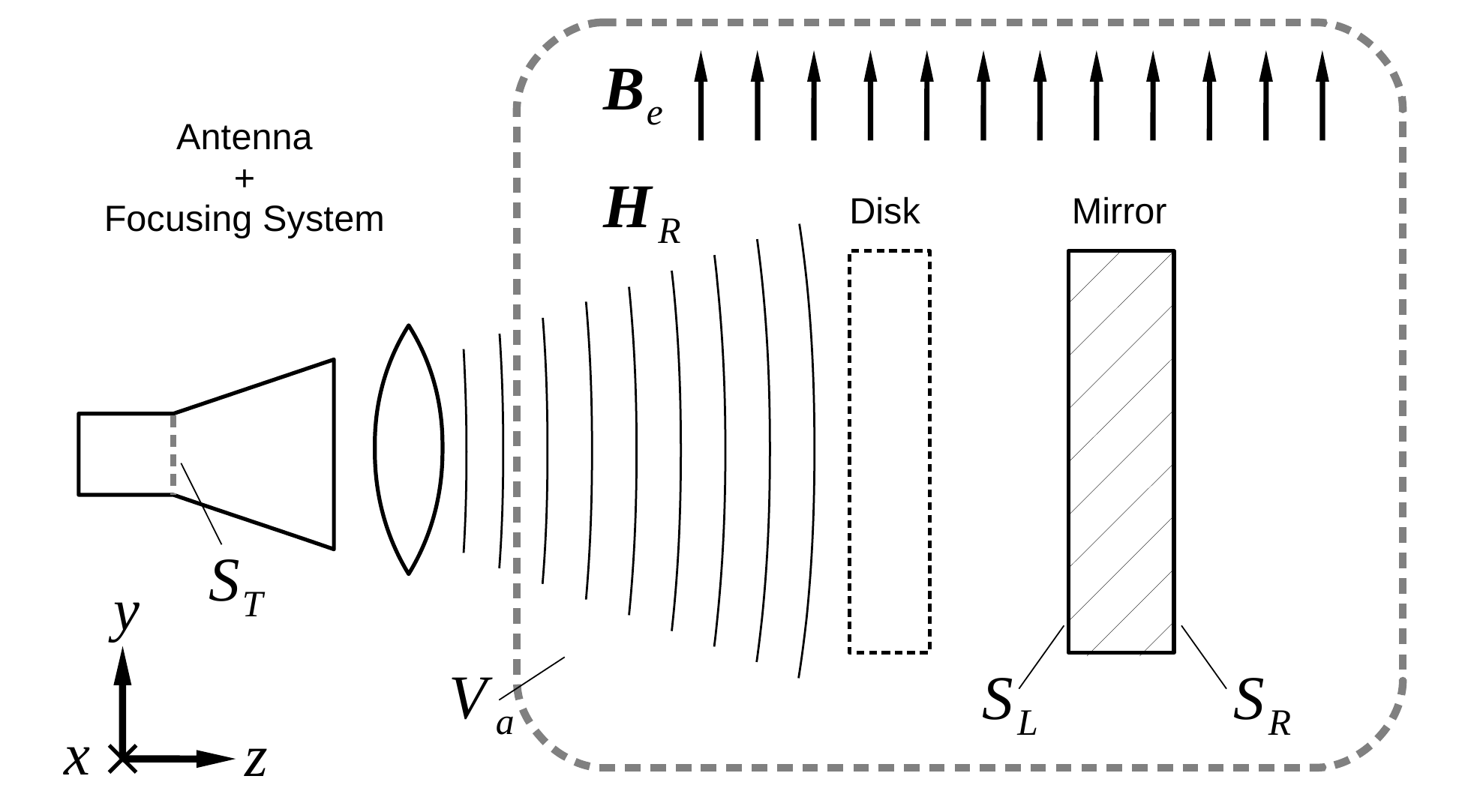}
    \caption{Sketch of a dish antenna (without disk) or dielectric haloscope (with disk). An antenna and possible focusing system induce magnetic field $\vb*{H}_R$ on the surface of mirror (and disks) that are magnetized in a homogeneous external magnetic field $\vb*{B}_e$.}
    \label{fig:dish_antenna}
\end{figure}

A dish antenna haloscope consists of a single metallic interface that when magnetized will emit axion-induced radiation \cite{horns2013}. In the simplest case, as shown in figure \ref{fig:dish_antenna}, the magnetized interface is a simple plane mirror with a homogeneous external magnetic field tangential to the mirror surface, in our case pointing in the y-direction. The axion-induced emissions from the interface have to be coupled to an antenna, possibly via a focusing system. We assume that the antenna and focusing system are not magnetized, i.e. are outside the conversion volume $V_a$. During a reflection measurement, the antenna and focusing system produce a beam propagating in the z-direction. The backside of the mirror should then be well isolated from the antenna beam and only the antenna facing side $S_L$ contributes to the axion signal power from equation \eqref{Eq:power_ax_surf_int}. From the interface conditions for a perfect electric conductor in a vacuum, 
\begin{equation}
 \vb*{n}\vdot\left[\left[\frac{\vb*{H}_R \cross  \vb*{B}_e}{\epsilon}\right]\right]=-\vb*{B}_e \vdot \vb*{j}_R,   
\end{equation}
where $\vb*{j}_R$ is the surface current density on the interface and we find
\begin{equation}
    P_{\mathrm{sig}} = \frac{g_{a\gamma}^2 \abs{a_0}^2\abs{\vb*{B}_e}^2}{16  P_{\mathrm{in}}}\abs{\int_{S_{L}} \dd{A} \vb*{\hat{y}}\vdot \vb*{j}_R}^2.
    \label{Eq:power_ax_dish_antenna}
\end{equation}
Irrespective of the details of the remaining, unmagnetized optical system, equation \eqref{Eq:power_ax_dish_antenna} tells us that we only need to know $\vb*{j}_R$ on the surface of the mirror. In particular, many effects that could happen along the optical axis such as losses, longitudinal standing waves, and coupling inefficiency from the antenna will be encoded in $\vb*{j}_R$ and do not have to be further characterized. In an idealized 1D limit without antenna effects, \eqref{Eq:power_ax_dish_antenna} agrees with previous calculations \cite{madmax_theo_found},
\begin{equation}
   P_{\mathrm{sig}} \xrightarrow{1D} \frac{g_{a\gamma}^2 \abs{a_0}^2 \abs{\vb*{B}_e}^2 A}{2} \equiv P_0,
\end{equation}
where $A$ is the area of the mirror. This can be seen by expanding $\vb*{H}_R$ into incoming and reflected plane waves. For an incoming wave with amplitude $H_{\mathrm{in}}$, the input power is $P_{\mathrm{in}}=1/2 A \abs{H_{\mathrm{in}}}^2$. At the mirror one finds $\abs{\vb*{j}_R}=2 \abs{H_{\mathrm{in}}}$, resulting in the above expression for $P_0$.


\subsection{Dielectric haloscope}
A dielectric haloscope adds dielectric disks to the dish antenna to further amplify $P_{\mathrm{sig}}$ \cite{caldwell2017}. 
From equation \eqref{Eq:power_ax_surf_int} we can see that the contribution of each interface needs to be added coherently. Since the dielectric disks are usually very thin compared to their diameter, we can neglect the contribution of the cylindrical shell. For a perfect dielectric with permittivity $\epsilon$ in a vacuum, one finds that
\begin{equation}
 \vb*{n}\vdot\left[\left[\frac{\vb*{H}_R \cross  \vb*{B}_e}{\epsilon}\right]\right]=\left(\frac{1}{\epsilon}-1\right)\vb*{B}_e \vdot \left(\vb*{n} \cross \vb*{H_R}\right),   
\end{equation}
The power boost factor $\beta ^2$ of a dielectric haloscope is defined as the ratio of total signal power $P_{\mathrm{sig}}$ to $P_0$, the idealized signal power of a single mirror. From equation \eqref{Eq:power_ax_surf_int} we find 
\begin{equation}
    \beta^2 = \frac{P_{\mathrm{sig}}}{P_0} = \frac{1}{2}\abs{\mathcal{P}_{\sigma} + \sum_i \left(\frac{1}{\epsilon_i}-1\right) \mathcal{P}_i}^2 ,
    \label{Eq:power_ax_dielectric_haloscope}
\end{equation}
where $\mathcal{P}_{\sigma}$ and $\mathcal{P}_i$ are the individual contributions of the plane mirror and dielectric disks, respectively:
\begin{equation}
\begin{aligned}
    \mathcal{P}_{\sigma} &= \frac{1}{2\sqrt{A P_{\mathrm{in}}}}\int_{S_{L}} \dd{A} \vb*{\hat{y}}\vdot \vb*{j}_R \\
    \mathcal{P}_{i} &= \frac{1}{2\sqrt{A P_{\mathrm{in}}}}\left(\int_{S_{R}} \dd{A} \vb*{\hat{x}}\vdot \vb*{H}_R - \int_{S_{L}} \dd{A} \vb*{\hat{x}}\vdot \vb*{H}_R \right).
\end{aligned}
\label{Eq:power_contrib}
\end{equation}
For the dielectric disks, both left and right faces contribute. In the so-called transparent mode where the disks become fully transparent for a phase depth of $\delta_{\epsilon}=\pi$, the left and right faces add up constructively. In the resonant mode, $\delta_{\epsilon}=\pi/2$, and the disks are maximally reflective. The left and right faces only partly interfere constructively, however, resonance effects increase the amplitude of $\vb*{H}_R$ relative to $P_{\mathrm{in}}$. Depending on the dielectric haloscope's mode of operation, one could further simplify the above expression for $\mathcal{P}_{i}$. Equation \eqref{Eq:power_ax_dielectric_haloscope} reduces in a 1D limit to previous results, namely the overlap integral approach of \cite{madmax_qft}. In this approach, derived from quantum field calculations, it is also the reflection-induced fields and not the axion-induced fields that enter the calculation for the signal power. 


\section{\label{sec:num_validation}Numerical Validation}
\begin{figure}
    \centering
    \includegraphics[width=\textwidth]{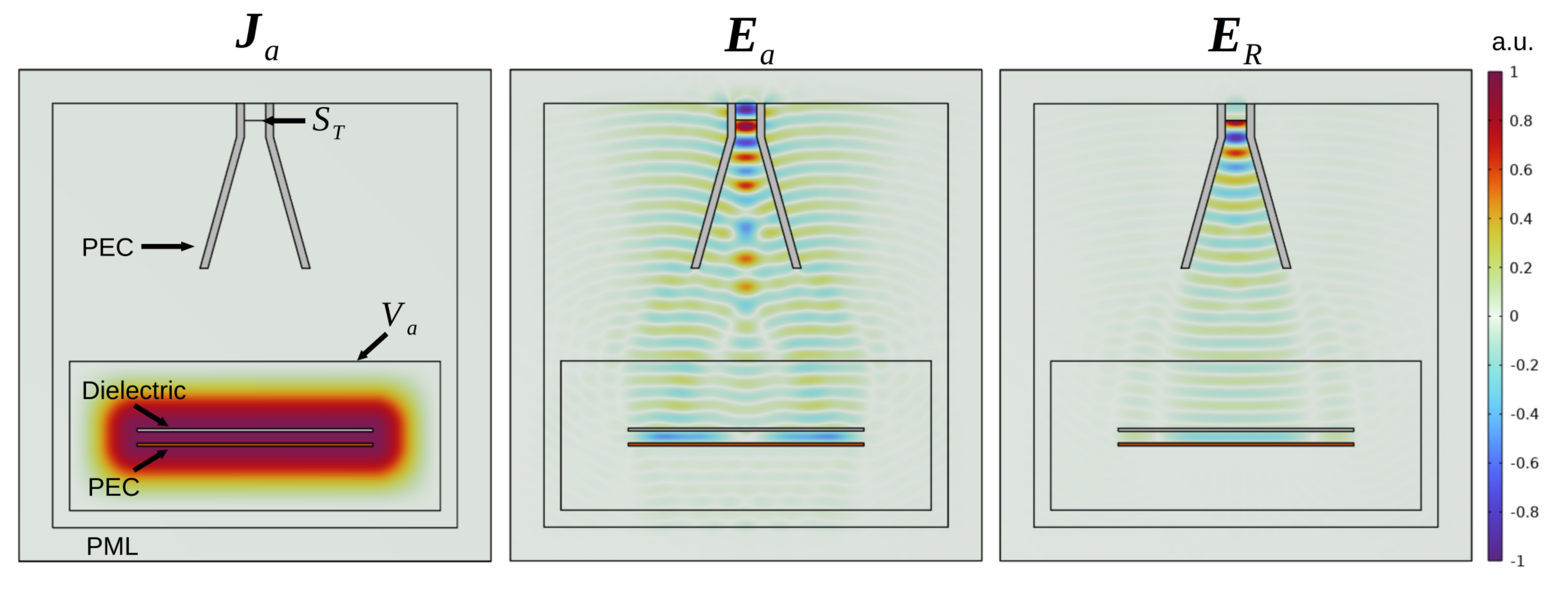}
    \caption{2D axisymmetric simulation of a minimal dielectric haloscope with a mirror, disk, and simple horn antenna. Left: axion effective current density $\vb*{J}_a$, polarized in the y-direction (pointing right) and confined to conversion volume $V_a$. Center: Resulting y-component of the axion-induced electric field. Right: Reflection-induced electric field from exciting a TE11 mode at $S_T$. All fields are normalized to their respective maxima.}
    \label{fig:comsol}
\end{figure}

\begin{figure}
    \centering
    \includegraphics[width=\textwidth]{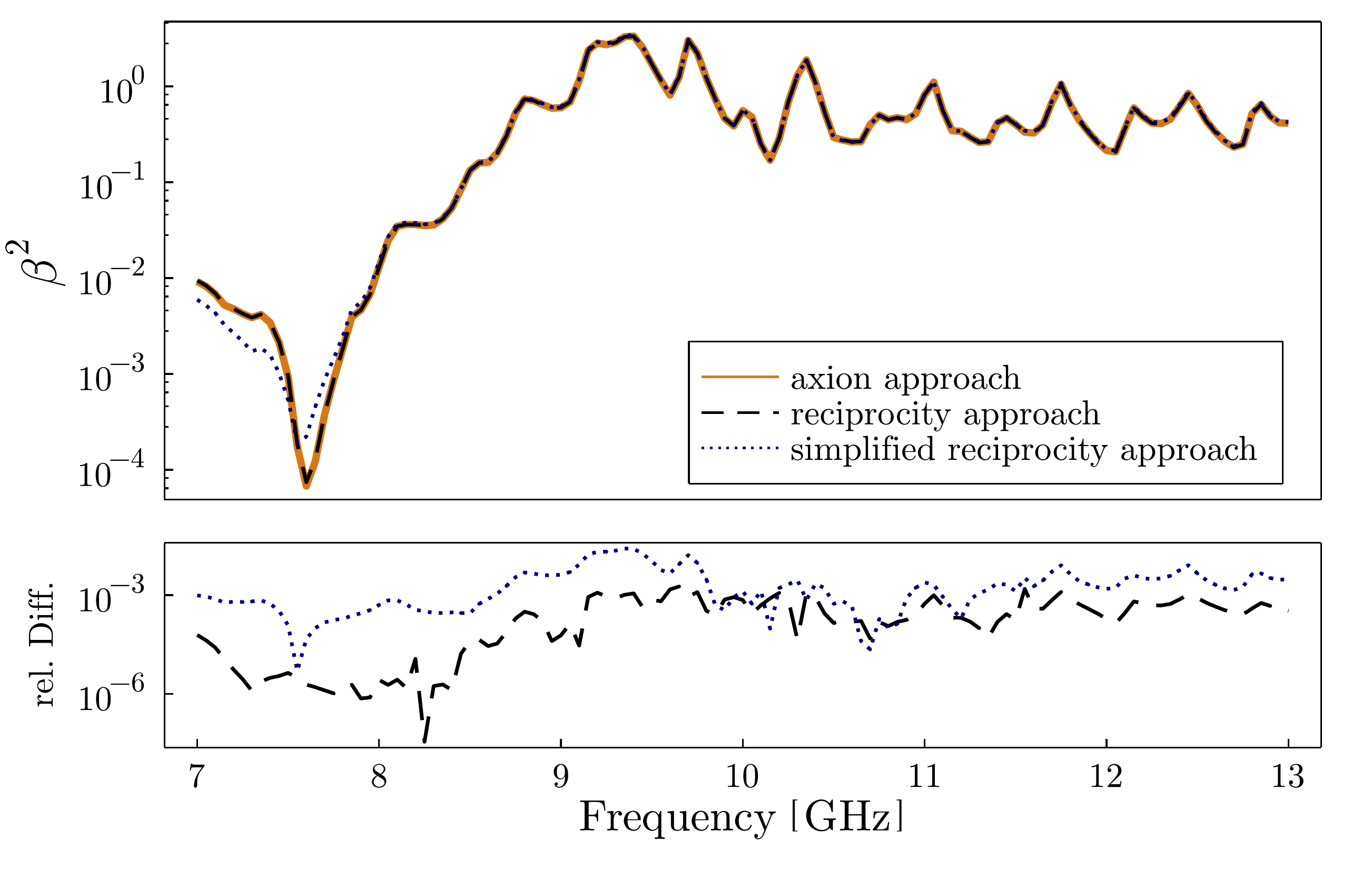}
    \caption{Comparison of power boost factor from axion and reciprocity approach for a minimal dielectric haloscope. In the axion approach (solid line), signal power is calculated from axion-induced fields via equation \eqref{Eq:power_ax_poynting}. In the reciprocity approach, signal power is calculated from the reflection-induced fields either using the main result of our paper, equation \eqref{Eq:power_ax_reciprocity} (dashed line), or the simplified version for a dielectric haloscope \eqref{Eq:power_ax_dielectric_haloscope} (dotted line). The relative difference, normalized to the maximum, between the axion and reciprocity approach is shown below. }
    \label{fig:reciprocity_vs_axion_approach}
\end{figure}

In this section, we validate the reciprocity approach with numerical simulations. As an example, the minimal dielectric haloscope of figure \ref{fig:dish_antenna} is simulated using COMSOL Multiphysics\textsuperscript{\tiny\textregistered}, a finite element simulation software. In this section, we will use SI units. The relative permittivity of the disk is that of sapphire, $\epsilon = 9.3$, while the mirror is a perfect electric conductor. The disk and mirror have a radius of $R=\SI{15}{\centi \metre}$ and a thickness of $d=\SI{3.69}{\milli \metre}$, corresponding to a phase depth of $3/4\pi$ at center frequency $f_c = \SI{10}{\giga \hertz}$ for the disk. The separation between disk and mirror is set to have a phase depth of $\pi$ at $f_c$. The antenna is a simple conical horn antenna made out of a perfect electric conductor. Its aperture radius is $2 \lambda_c$ where $\lambda_c = c/f_c$. Its axial length is $7 \lambda_c$. It is connected to a cylindrical waveguide with a radius of $\SI{1.35}{\centi\metre}$ and length of $\SI{4.28}{\centi \metre}$. At the chosen frequency range it supports only the TE11 mode. A perfectly matched layer surrounds the entire simulation domain. To simulate linearly polarized fields in an axisymmetric geometry, we decompose all relevant field components $X$ into a left and right circularly polarized part \cite{Knirck_2019}:
\begin{equation}
    X(r,\phi,z) =  X^+(r,z)e^{-i \phi} +  X^-(r,z)e^{+i \phi}
    \label{Eq:azimuthal_decomp}
\end{equation}
Only the azimuthally symmetric components $X^{\pm}(r,z)$ are simulated and later added according to \eqref{Eq:azimuthal_decomp}. The two following scenarios are simulated: in the axion approach, we set an external current density acting as $\vb*{J}_a$ in the conversion volume $V_a$. It points in the y-direction and has a constant value over disk and mirror before slowly decreasing to zero, as shown in the left panel of figure \ref{fig:comsol}. We then simulate the resulting axion-induced fields (center panel of figure \ref{fig:comsol}) and calculate the axion signal power via equation \eqref{Eq:power_ax_poynting}. 
In the reciprocity approach, we use a port boundary condition to excite an outgoing TE11 mode at $S_T$. This excites the reflection-induced fields, shown in the right panel of figure \ref{fig:comsol}, and we then use equation \eqref{Eq:power_ax_reciprocity} to calculate again the axion signal power. A comparison between these two approaches is shown in figure \ref{fig:reciprocity_vs_axion_approach}. The two approaches match almost perfectly with deviations less than $\SI{1.3}{\percent}$ relative to the maximum. In addition, we also show the axion signal power according to equation \eqref{Eq:power_ax_dielectric_haloscope}, the simplified version for a dielectric haloscope. As expected, the deviation is slightly larger, less than $\SI{2.3}{\percent}$, since we are essentially ignoring contributions from the small but nonzero $\curl{\vb*{J}_a}$. The setup simulated here is far from ideal as clearly there is a poor coupling between the antenna and axion-induced fields. It serves, however, as an illustrative example of the power of the reciprocity approach even when axion-induced and reflection-induced fields are very different. 

Another advantage of the reciprocity approach is that simulating the reflection-induced fields also yields experimental parameters like the reflection coefficient, which cannot be calculated from the axion-induced fields. One can thus entirely skip the axion approach, effectively halving the computational cost of simulating both axion-induced power and desired experimental parameters.\footnote{For the setup shown here, both reciprocity and axion approach need about $\SI{9}{}-\SI{12}{\second}$ for a solution at a single frequency running on two AMD EPYC 7543 32-Core processors. Evaluating the integrals of equation \eqref{Eq:power_ax_reciprocity} and \eqref{Eq:power_ax_dielectric_haloscope} takes $<\SI{0.1}{\second}$.} Moreover, once $\vb*{E}_R$ has been calculated, one can input different $\vb*{J}_a$ into equation \eqref{Eq:power_ax_reciprocity} with almost no additional computational cost beside evaluating the integral. This is particularly useful for investigating the effect of non-zero axion velocity or external field inhomogeneities which both change $\vb*{J}_a$ \cite{Knirck_2018}.


\section{\label{sec:exp_consideration}Experimental considerations}
So far we have shown that it is in principle possible to calculate the expected axion signal power purely from directly measurable quantities and axion parameters. How one can achieve this experimentally is often a different story. The three quantities we need to measure are the external fields $\vb*{B}_e$ or $\vb*{E}_e$, the reflection-induced electric field $\vb*{E}_R$ and the input power $P_{\mathrm{in}}$. Knowing the external fields is already a requirement for any haloscope design and should not pose an additional challenge. $P_{\mathrm{in}}$ and $\vb*{E}_R$ are likely to be more challenging. How to exactly measure these will depend on the nature of the setup. A generally valid and promising approach is non-resonant perturbation theory \cite{steele1966}. Its resonant formulation, also known as cavity perturbation theory or Slater's perturbation theory \cite{slater46,slater1952}, is widely applied in the so-called bead pull method \cite{rapidis2018}. The non-resonant formulation, however, also applies to open and broadband systems. It relates the local fields at a perturbing object inside the optical system to changes in the complex reflection coefficient $\Gamma$. For a small dielectric object, the change in $\Gamma$ is  
\begin{equation}
\Delta \Gamma = \frac{\alpha_e \omega}{P_{\mathrm{in}}} \vb*{E}_{R}^2 . 
\label{Eq:nonresonant_small_steele}
\end{equation}
where $\alpha_e$ depends on the permittivity and geometry of the object. One immediate advantage of equation \eqref{Eq:nonresonant_small_steele} is that if we constrain $\vb*{E}_R$ from \eqref{Eq:nonresonant_small_steele} and insert into \eqref{Eq:power_ax_reciprocity}, the input power $P_{\mathrm{in}}$ cancels and we have one quantity less to worry about. Contrast this to measuring $\vb*{E}_R$ with fields probes: here the measured signal will be proportional to $P_{\mathrm{in}}$ requiring an additional measurement on $P_{\mathrm{in}}$. Instead of introducing a new object into the optical system, one could also shift or deform existing components by small amounts. Many haloscopes already tune their resonant frequency this way for example with tuning rods in cavities or disks in a dielectric haloscope.
Measuring the reflection-induced fields in an open dielectric haloscope by both precise disk movement and the bead-pull method is actively being investigated and will be the subject of future publications.


\section{Conclusion}
We have used reciprocity relations to derive a general expression for the axion signal power of a haloscope. It is applicable to a variety of different designs, from (multiple) cavity approaches to dielectric haloscopes, and provides a unifying description of formerly distinct setups. More importantly, however, it allows for predicting the expected axion signal power from measurable quantities and axion parameters alone. Specifically, this expression does not depend on the axion-induced electromagnetic fields that are fundamentally unknown. Instead, it depends on the reflection-induced electromagnetic fields which are in principle measurable. We applied this new approach to arbitrarily shaped interfaces in a magnetic field and then simplified it further to a dielectric haloscope. As limiting cases, our approach agrees with previous results for cavities, dish antennas, and dielectric haloscopes. Furthermore, we validated our results with numerical simulations. On the experimental side, non-resonant perturbation theory seems to be a promising approach to constrain the reflection-induced fields.


\begin{acknowledgments}
The author would like to thank the MADMAX collaboration and in particular its simulation group for helpful discussions. Special thanks go to Stefan Knirck, Alexander Miller, Javier Redondo, Frank Steffen, and Erika Garutti for their inspiration, feedback, and support. This work is supported by the Deutsche Forschungsgemeinschaft (DFG, German Research Foundation) under Germany’s Excellence Strategy, EXC 2121, Quantum Universe (390833306).
\end{acknowledgments}


\appendix
\section{\label{app:cavity_power}Cavity haloscope signal power}

The axion signal power from equation \eqref{Eq:power_ax_reciprocity} reduces to the well-established expression cavity haloscopes use for the axion signal power. We assume a homogeneous external magnetic field and that all of the optical system is magnetized. For simplicity, we also assume that the resonance frequency of the cavity matches the axion frequency $\omega_a$. We start by inserting the form factor defined as 
\begin{equation}
    C= \frac{\abs{\int_{V_a} \dd{V} \vb*{E}_R \vdot \vb*{B}_e}^2}{V_a \abs{B_e}^2 \int_{V_a} \dd{V} \abs{\vb*{E}_R}^2}
\end{equation}
into equation \eqref{Eq:power_ax_reciprocity}. With the average energy stored in the cavity
\begin{equation}
    U = \frac{1}{2}\int_{V_a} \dd{V} \abs{\vb*{E}_R}^2,
\end{equation}
one finds that
\begin{equation}
    P_{\mathrm{sig}} = \frac{\omega_a^2 g_{a\gamma}^2 \abs{a_0}^2 }{8 P_{\mathrm{in}}} C V_a \abs{B_e}^2 U .
    \label{Eq:power_ax_reciprocity_cavity_1}
\end{equation}
For a cavity coupled to a transmission line, the ratio of stored energy $U$ and input power $P_{\mathrm{in}}$ can be expressed in terms of loaded quality factor $Q_L$ and coupling coefficient $\kappa$ as 
\begin{equation}
    \frac{\omega_a U}{P_{\mathrm{in}}} = \frac{4 \kappa}{\kappa+1} Q_L. 
\end{equation}
Together with $\rho_a = m_a^2 \abs{a_0}^2 / 2$, this then yields
\begin{equation}
     P_{\mathrm{sig}} = \frac{g_{a \gamma} ^2 \rho_a \omega_a}{m_a^2}   \frac{\kappa}{1+\kappa}  Q_L V_a B_e^2 C   
     \label{Eq:cavity_power_ax_A}
 \end{equation}
which agrees with literature \cite{admx_analysis2021,capp2021}.

\bibliographystyle{unsrt}
\bibliography{bibliography}

\end{document}